\documentstyle[titlepage,12pt]{article}

\def\sgn{\mbox{sgn}}
\def\be{\begin{equation}}
\def\bea{\begin{eqnarray}}
\def\tphi{\tilde{\phi}}
\def\tpsi{\tilde{\psi}}
\def\trho{\tilde{\rho}}
\def\grad{\nabla}
\def\ee{\end{equation}}
\def\eea{\end{eqnarray}}

\def\dif{\partial}

\newcount\sectionnumber
\def\sect
{\global\equationnumber=0
\global\advance\sectionnumber by 1
\the\sectionnumber . }
\newcount\equationnumber
\def   \num
{\eqno{\global\advance\equationnumber by 1
\left(\the\sectionnumber .\the\equationnumber \right)}}

\begin{document}
\title{Liouville Black Holes}
\author{R.B. Mann \\
Department of Physics \\
University of Waterloo \\
Waterloo, Ontario \\
N2L 3G1}

\date{August 5, 1993\\
WATPHYS-TH93/03}

\maketitle

\begin{abstract}

The dynamics of Liouville fields coupled to gravity are investigated by
applying the principle of general covariance to the Liouville action in the
context of a particular form of two-dimensional dilaton gravity. The
resultant field equations form a closed system for the Liouville/gravity
interaction. A large class of  asymptotically flat solutions to the field
equations is obtained, many of which can be interpreted as black hole
solutions. The temperature of such black  holes is proportional to their
mass-parameters. An exact solution to the back reaction problem is obtained
to one-loop order, both for conformally coupled matter fields and for the
quantized metric/Liouville system. Quantum effects are shown to map the
space of classical solutions into one another. A scenario for the end-point
of black-hole radiation is discussed.

\end{abstract}

\section{Introduction}

Two-dimensional gravity continues to be a subject of intense study,
motivated both by string theory \cite{Polya}  and by a desire to study
quantum gravitational effects in a mathematically tractable setting
\cite{Odintsov}.   Black hole solutions \cite{BHT,MST,MFound,EDW} have
especially  attracted attention as the simplicity of $(1+1)$ dimensions
permits  significant clarification of the conceptual issues associated with
black hole radiation.

In a spacetime of two dimensions, the construction of a  gravitational
theory necessarily departs from the Einstein-Hilbert  formalism because of
the triviality of the Einstein tensor:  $G_{\mu\nu}[g]\equiv 0$ for all
metrics $g_{\mu\nu}$. As a consequence,  the Einstein-Hilbert action is
trivial, and new approaches must be taken.  These have included non-local
actions \cite{Polya}, higher-derivative theories \cite{Higher}, setting the
Ricci scalar equal to a constant \cite{JackTeit}, setting the one-loop beta
functions of the bosonic non-linear sigma model to zero in a
two-dimensional  target space \cite{MSW}, or setting the Ricci scalar equal
to the trace of the 2D stress-energy tensor \cite{MST,MFound}. In each
approach the Ricci scalar becomes a non-vanishing function of the
co-ordinates over some region of spacetime, permitting the spacetime to
develop interesting features such as black-hole horizons and
singularities.

The study of Liouville fields has been of particular interest, since the
non-local action of ref. \cite{Polya} becomes the (local) Liouville action
for a certain choice of co-ordinates.  In this approach, the conformal
factor of the metric is the Liouville field. This has led most theorists to
adopt the general paradigm of interpreting the Liouville action as the
action for 2D (quantum) gravity \cite{DDK}.

In the present paper a different approach towards the relationship between
the Liouville action and 2D gravity is developed.
Rather than obtaining the Liouville action as a form of induced
gravity \cite{Polya} which arises from quantum effects (and in which the
Liouville field is identified  with the conformal factor of the metric
\cite{DDK}),
the approach taken here is to consider an independent Liouville field  in a
curved $(1+1)$-dimensional spacetime in such a way that its stress-energy
generates gravity. This may be done by taking the (flat-space) Liouville
action and applying to it the principle of general covariance.
The net effect of this is to set the Ricci scalar equal to
the trace of the Liouville stress-energy, an approach which has been widely
studied for other forms of two-dimensional stress-energy
in a variety of previous contexts \cite{MFound,RGRG}.

More specifically, exact solutions to the field equations which follow from
the action of a Liouville field in a curved 2D spacetime are derived and
their consequences investigated, extending the work of ref.
\cite{Qlou}. The most general form of the action which is of lowest
order in powers of the curvature depends upon three independent
coupling constants; a choice of these three constants is
equivalent to a choice of theory.  An exact solution within a
given theory then yields a spacetime  which is a function  of
these parameters (along with the constants of integration that
specify the particular solution). Conditions necessary for the
existence of asymptotically flat black hole solutions of positive
mass constrain the space of coupling constants, and it is shown
that a wide class of theories exist which permit such solutions.
One of these solutions \cite{RtDil} closely resembles that of the
string-theoretic 2D black hole \cite{EDW,MSW}. However the
thermodynamic properties of this (and the other) black hole
solutions markedly differ from those of the string-theoretic
case: the temperature is proportional to the ADM-mass parameter
(as opposed to being a constant), and the entropy varies
logarithmically with the mass.

Quantum corrections due to the presence of conformally coupled matter  are
then taken into account, and an exact solution to the field equations which
includes the back-reaction is obtained. The effects of the   back-reaction
are to map the space of solutions into itself, so that a (classical)
solution with a given set of parameters is mapped into one with another set
of parameters that may or may not satisfy the criteria for the existence of
asymptotically flat black holes of positive mass. When quantum effects due
to the gravitational and Liouville fields are taken into account, a similar
phenomenon happens, except that in this case the map of the space of
solutions into itself is somewhat more complicated.

The paper is organized as follows. In section 2 the general form of the
aciton is described and the field equations are derived. In section 3 the
general form of the exact solutions is obtained, and in section 4 the
criteria necessary for black hole solutions is discussed and applied to the
solutions obtained in section 3. Section 5 is devoted to a consideration of
the thermodynamic properties Liouville black holes, and section 6 to the
incorporation of quantum effects. Section 7 contains some concluding
remarks.

\section{Two-Dimensional Gravity Coupled to a Liouville Field}

The action considered here is taken to be \cite{RtDil,semi}
\be
S = S_G + S_M = \int d^2x\sqrt{-g}\left(
\frac{1}{8\pi G}[\frac{1}{2} g^{\mu\nu}\grad_\mu\psi
   \grad_\nu\psi +\psi R] + {\cal L}_M \right) \label{1}
\ee
where $R$ is the Ricci scalar and $ S_M = \int d^2x\sqrt{-g}{\cal L}_M$ is
the two dimensional matter action which is independent of the auxiliary
field $\psi$. The action (\ref{1}) is designed so that the auxiliary field
$\psi$ has no effect on the classical
evolution of the gravity/matter system. This
is easily seen by considering the field equations:
\be
\grad^2\psi - R = 0  \label{2}
\ee
\begin{equation}
\frac{1}{2}\left(\grad_\mu\psi\grad_\nu\psi - \frac{1}{2} g_{\mu\nu}
(\grad\psi)^2\right) +
g_{\mu\nu}\grad^2\psi - \grad_\mu\grad_\nu\psi
= 8\pi G T_{\mu\nu} \label{3}
\end{equation}
where $T_{\mu\nu}$ is the stress-energy tensor of two-dimensional matter;
it is easily seen to be conserved by taking the divergence of (\ref{3}).
Insertion of the trace of (\ref{3}) into (\ref{1}) yields
\be
R = 8\pi G T_\mu^{\ \mu}  \qquad \mbox{where}\qquad
\grad_\nu T^{\mu\nu} = 0 \label{2a}
\ee
the evolution of $\psi$ then being determined only by the traceless part of
(\ref{3}).

The theory associated with the action (\ref{1}) is of interest
because it yields a two dimensional theory of gravity which closely
resembles $(3+1)$ dimensional general relativity in that the (classical)
evolution of the gravitational field is driven by the stress-energy and no
other Brans-Dicke type fields \cite{semi}. Its classical and semi-classical
properties and solutions are also markedly similar
\cite{MST,semi,Arnold,Shardir,TomRobb}. Indeed, the field equations which
follow from (\ref{18}) may be obtainted from a reduction of the Einstein
equations from $D$ to 2 spacetime dimensions \cite{Rossred}.

Incorporation of the Liouville action
\be
S_L = \int d^2x [b (\grad\phi)^2 + \Lambda e^{-2a\phi}] \label{4}
\end{equation}
is easily carried out by applying the principle of general covariance, so
that the action which couples the Liouville field $\phi$ to gravity is
given by
\be
S = S_G + S_L = S_G + \int d^2x \sqrt{-g}[b (\grad\phi)^2
+ \Lambda e^{-2a\phi} + \gamma \phi R] \label{5}
\end{equation}
the Liouville action being viewed as a particular form of the matter
action $S_M$. In (\ref{5})
a non-minimal coupling $\gamma \phi R$ has been included (for a
canonically normalized $\phi$, $b=-\frac{1}{2}$).   Note that in other
approaches to Liouville-gravity \cite{DDK} $S_G=0$; the Liouville field in
(\ref{5}) is then interpreted as the conformal factor associated with the
metric so that $g_{\mu\nu}= e^{\phi}\hat{g}_{\mu\nu}$, $\hat{g}_{\mu\nu}$
being a fixed background metric with respect to which the action is not
varied. Here however, the Liouville field $\phi$ is considered
{\it a-priori} to be an independent matter field coupled to gravity,  so
that the action in (5) is a functional of $\psi$,  $g_{\mu\nu}$ and $\phi$
($S=S[\psi,g,\phi]$).  This action  depends upon only three independent
coupling constants, since one of $a$, $b$, or $\gamma$ may be absorbed by
rescaling $\phi$.  However in order to more easily take into account
possible overall sign changes in each of the terms, all four coupling
constants will be retained.

The field equations which follow from (\ref{5}) are
\be
\grad^2\psi - R = 0  \label{6}
\ee
\be
-2b\grad^2\phi +\gamma R -2a \Lambda e^{-2a\phi}= 0  \label{7}
\ee
and
\begin{eqnarray}
\frac{1}{2} {\left( \grad_\mu\psi\grad_\nu\psi - \frac{1}{2} g_{\mu\nu}
(\grad\psi)^2 \right)} &+&
g_{\mu\nu} \grad^2\psi - \grad_\mu\grad_\nu\psi \qquad \nonumber \\
\quad = 8\pi G \left[ -b \left( \grad_\mu\phi\grad_\nu\phi
- \frac{1}{2} g_{\mu\nu} (\grad\phi)^2 \right) \right.
&-& \left. \gamma \left( g_{\mu\nu}\grad^2\phi
+\grad_\mu\grad_\nu\phi \right)  \right. \nonumber \\
\left. \quad +\frac{1}{2} g_{\mu\nu}\Lambda e^{-2a\phi} \right] \label{8}
\end{eqnarray}
It is easily checked that the left-hand-side of (\ref{8}) is
divergenceless when (\ref{6}) holds and that the right-hand-side is
divergenceless when (\ref{7}) holds. Hence the system (\ref{6}--\ref{8})
is a system of 5 equations with two identities, allowing one to solve for
the three unknown fields $\psi$, $\phi$ and the one independent degree of
freedom in the metric.

Making use of the trace of (\ref{8}), the field equations (\ref{6}) and
(\ref{7}) become
\be
(b + 4\pi G\gamma^2)\grad^2\phi = (4\pi G\gamma -a)\Lambda e^{-2a\phi}
\label{9}
\ee
\be
(b + 4\pi G\gamma^2) R = 8\pi G\Lambda (a\gamma + b) e^{-2a\phi} \label{10}
\ee
with the evolution of $\psi$ being determined by the traceless part of
(\ref{8}).

In conformal coordinates
\be
g_{+-} = -\frac{1}{2}e^{2\rho} \qquad R= 8 e^{-2\rho}\dif_{+}\dif_{-}\rho
\qquad \grad^2\phi = -4e^{-2\rho}\dif_{+}\dif_{-}\phi
\label{11}
\ee
the field equations simplify to
\be
\dif_{+}\dif_{-}(\psi + 2\rho) = 0            \label{12}
\ee
\be
\dif_{+}\dif_{-}\rho = \frac{\pi G\Lambda(a\gamma+b)}{b+4\pi G\gamma^2}
e^{2(\rho-a\phi)} \label{13}
\ee
\be
\dif_{+}\dif_{-}\phi = \frac{\Lambda(a-4\pi G\gamma)}{4(b+4\pi G\gamma^2)}
e^{2(\rho-a\phi)} \label{14}
\ee
\be
\frac{1}{2}(\dif_{\pm}\psi)^2 -\dif^2_\pm\psi
+ 2 \dif_\pm\rho\dif_\pm\psi
+ 8\pi G(b(\dif_{\pm}\phi)^2 -\gamma\dif^2_\pm\phi
+ 2\gamma \dif_\pm\rho\dif_\pm\phi)  = 0   \label{15}
\ee
where the last set of equations follows from the traceless part of (\ref{8}).

\section{Exact Solutions}

Provided that $b\neq -4\pi G\gamma^2$ the field equations (\ref{13}) and
(\ref{14}) are non-degenerate. Taking an appropriate linear combination of
these two equations yields
\be
\dif_{+}\dif_{-}(\rho - a\phi) =
\frac{8\pi G a\gamma + 4\pi Gb - a^2}{4\pi G(a\gamma+b)}
e^{2(\rho-a\phi)}  \label{16a}
\ee
{\it i.e.} $\rho - a\phi$ obeys the Liouville equation, provided
$b\neq -4\pi G\gamma^2$.

The solution to (\ref{16a}) is well-known
\be
\rho-a\phi = \frac{1}{2}
\ln\left[\frac{f^\prime_{+}f^\prime_{-}}{(\frac{A}{K^2}-K^2f_{+}f_{-})^2}
\right]                 \label{18}
\ee
where $K$ is a constant of integration and $f_\pm$ are arbitrary functions of
the coordinates
$x_\pm$ and the prime refers to differentiation by the relevant functional
argument. The constant $A$ is
\be
A = \Lambda\frac{4\pi G b +8\pi G a\gamma-a^2}{4(4\pi G\gamma^2+b)}
e^{-2a\phi_0}
\label{19}
\ee
where from (\ref{14})
\be
\phi = \frac{a-4\pi G\gamma}{4\pi G(a\gamma+b)}\rho  +
\frac{1}{a}(h_{+} + h_{-}) +\phi_0 \label{16b}
\ee
(provided $a\gamma+b\neq 0$) where $h_\pm$ are
arbitrary functions of the coordinates $x_\pm$ respectively; for
convenience an additional constant of integration $\phi_0$ has been
retained. From (\ref{13})
\be
\rho = \xi
\ln \left[\frac{f^\prime_{+}f^\prime_{-}}{(\frac{A}{K^2}-K^2f_{+}f_{-})^2}
\right] + h_+ + h_-
\label{17}
\ee
where
\be
\xi = \frac{2\pi G(a\gamma+b)}{4\pi Gb+8\pi Ga\gamma - a^2}
\label{17a}
\ee
yielding a metric
\begin{eqnarray}
ds^2 &=& - e^{2\rho}dx_+ dx_-
= -\left[\frac{f^\prime_{+}f^\prime_{-}}{(\frac{A}{K^2}-K^2f_{+}f_{-})^2}
\right]^{2\xi} e^{2(h_+ + h_-)}dx_+ dx_-  \nonumber \\
&=&
 -\frac{df_+ df_- }{(\frac{A}{K^2}-K^2f_{+}f_{-})^{4\xi}}
\label{23}
\end{eqnarray}
where the last term follows from a suitable choice of $h_\pm$.

Equation (\ref{12}) has the solution
\be
\psi = -2\rho +\chi_+(x_+) + \chi_-(x_-) \label{17b}
\ee
and equations (\ref{15}) constrain these functions to obey
\be
\frac{1}{8\pi G}\left(\frac{1}{2}(\chi^\prime_\pm)^2 -
\chi^{\prime\prime}_\pm\right) -
\frac{b+4\pi G\gamma^2}{2(4\pi Gb+8\pi Ga\gamma - a^2)}
\left[\left(\frac{f^{\prime\prime}_\pm}{f^\prime_\pm}\right)^\prime
- \frac{1}{2}\left(\frac{f^{\prime\prime}_\pm}{f^\prime_\pm}\right)^2
\right]= 0
\label{20}
\ee
which has the solution
\be
\chi_\pm(x_\pm) = \ln\left[
f_\pm^\prime\left(\sqrt{
\frac{4\pi G(b+4\pi G\gamma^2)}{4\pi Gb+8\pi Ga\gamma - a^2}}
x_\pm\right)\right] \qquad .\label{20a}
\ee

If $A=0$ the solution (\ref{16a}) becomes
\be
a\phi = \rho + \frac{1}{2}\ln(\hat{k}^2 g_+^\prime g_-^\prime) +\phi_0
\label{26a}
\ee
where $g_\pm$ are arbitrary functions of the coordinates $x_\pm$
respectively. From (\ref{13}) and (\ref{14})
\be
\rho = -D\hat{k}^2g_+ g_- + h_+ + h_-                  \label{26}
\ee
where $h_\pm$ are again arbitrary functions of $x_\pm$ respectively
and
\be
D = \pi G\Lambda\frac{a}{4\pi G \gamma - a}  \label{27}
\ee
yielding a metric
\be
ds^2 = e^{2(h_+ + h_- - D\hat{k}^2 g_+ g_-)}dx_+ dx_-
= e^{-2D\hat{k}^2 g_+ g_-}dg_+ dg_-
  \label{28}
\ee
where again the last relation follows by making a suitable choice of
$h_\pm$. The analogue of (\ref{20}) for this choice is
\be
\frac{1}{8\pi G}\left(\frac{1}{2}(\chi^\prime_\pm)^2 -
\chi^{\prime\prime}_\pm\right)
+\left(\frac{g^{\prime\prime}_\pm}{g^\prime_\pm}\right)^\prime
- \frac{1}{2}\left(\frac{g^{\prime\prime}_\pm}{g^\prime_\pm}\right)^2
= 0
\label{20z}
\ee
with
\be
\chi_\pm(x_\pm) = \ln\left[
g_\pm^\prime\left(\sqrt{4\pi G}x_\pm\right) \label{20y} \right]
\ee
being the solution.

If $a\gamma+b=0$ then the spacetime is flat, $\psi$ is a free scalar
field and $\phi$ obeys the flat-space Liouville equation. Alternatively, if
$a=4\pi G\gamma$, then $\phi$ is a free scalar field and the spacetime has
constant curvature.

Each of the metrics (\ref{23},\ref{28}) may be transformed
to a static system of coordinates
\be
ds^2 = -\alpha(x) dt^2 + \frac{dx^2}{\alpha(x)}            \label{24}
\ee
under an inverse Kruskal-Szekeres transformation. There are three distinct
classes of solutions.
\bigskip

\noindent
({\bf A}) $\xi\neq 1/4$, $A\neq 0$

In this case the solution (\ref{23}) is
\begin{eqnarray}
\alpha(x) &=& 2Mx -  \frac{A}{M^2(1-p)^2}(2Mx)^p
= 2Mx -  B x_0^2 (\frac{x}{x_0})^p  \nonumber\\
\phi(x) &=&  \frac{2-p}{2a}\ln(2Mx) +\phi_0 =
\frac{2-p}{2a}\ln(x/x_0)   \label{25}\\
\psi(x) &=&  -p \ln(2Mx) +\psi_0 \nonumber
\end{eqnarray}
where  $K= M/(p-1)$ and
\be
p = \frac{4\xi}{4\xi -1} = 8\pi G \frac{a\gamma+b}{a^2+4\pi Gb} \label{25a}
\ee
\be
B = \frac{4A}{(p-1)^2} = \Lambda
\frac{(a^2+4\pi G b)^2 }{(b+4\pi G\gamma^2)
(4\pi Gb+8\pi Ga\gamma - a^2)}
\label{26b}
\ee
with $\phi_0 = \frac{p-2}{2a}\ln(2M x_0)$.
\bigskip

\noindent
({\bf B}) $\xi = 1/4$, $A\neq 0$

In this case (\ref{23}) is
\begin{eqnarray}
\alpha(x) &=& 1- C e^{-2M(x-x_0)} \nonumber\\
\phi(x) &=&    \frac{M}{a}(x-x_0)  \label{29}\\
\psi(x) &=&  2Mx +\psi_0 \nonumber
\end{eqnarray}
where now
\be
b = -\frac{a^2}{4\pi G}   \qquad
C \equiv \frac{2\pi G\Lambda a}{M^2(a+ 4\pi G\gamma)}
 \label{29b}
\ee
and $K=M$.
\bigskip

\noindent
({\bf C}) $A = 0$

In this case $p=1$ or
\be
b = \frac{a^2}{4\pi G} - 2a\gamma  \label{29d}
\ee
and
\begin{eqnarray}
\alpha(x) &=& 2Mx\ln(2Dx x_0)  \nonumber\\
\phi(x) &=&   \frac{1}{2a}\ln(\frac{4D}{M}x)  \label{30}\\
\psi(x) &=&  -2M\ln(\frac{D}{M}x) +\psi_0 \nonumber
\end{eqnarray}
where $D$ is given by (\ref{27}) and $\hat{k}^2=x_0/2$.

In each of (\ref{25}), (\ref{29}) and (\ref{30})
$\psi_0$, $M$ and $x_0$ are constants of integration, with $M$ being
proportional to $K$  for solution ({\bf A}) and $x_0$ replacing the
constant $\phi_0$ in (\ref{16b},{\ref{26a}).

The solutions (\ref{23},\ref{28}) (or alternatively
(\ref{25},\ref{29},\ref{30}) ) are, up to coordinate transformations,
the most general set of solutions to the field equations for arbitrary
$a$, $b$, $\gamma$ and $\Lambda$ in the absence of any other forms of
stress-energy.  The only restriction on the parameters is that
$b\neq -4\pi G\gamma^2$; otherwise the system (\ref{9},\ref{10}) becomes
underdetermined.  Note that none of (\ref{25},\ref{29},\ref{30}) are
invariant under $x\to -x$. As such they may be matched onto a
solution of collapsing matter which is moving either rightward or leftward.
Alternatively,  one may consider these
solutions as exterior to a localized system of two-dimensional matter by
making the replacement $2Mx \to 2M_{\pm}|x| - X_{\pm}$, where the `$+$"
refers to the rightward side of the localized stress-energy and the `$-$'
to the leftward side. If the localized stress-energy is centered about the
origin, then $\pm$ refers to positive/negative $x$.

Consider next the $M\to 0$ limit of (\ref{25},\ref{29},\ref{30}).  It will
be shown in the next section that $M$ is proportional to the ADM-mass. For
(\ref{25}) this limit is straightforward:
\begin{eqnarray}
ds^2 &=&
B x_0^2 (\frac{x}{x_0})^p dt^2 + \frac{dx^2}{-B x_0^2 (\frac{x}{x_0})^p}
\nonumber\\
\phi(x) &=&  \frac{2-p}{2a}\ln(x/x_0)   \label{25z}\\
\psi(x) &=&  -p \ln(x/{x}_0) +\hat{\psi}_0 \nonumber
\end{eqnarray}
(where $\psi_0$ has been shifted) and  corresponds to
\begin{eqnarray}
ds^2 &=&
-4x_0^2 B \left[\frac{Bx_0}{4\xi-1}(x_+ + x_-)\right]^{-4\xi} dx_+ dx_-
\nonumber\\
\phi(x) &=&  \frac{1-2\xi}{a}\ln(\frac{A x_0}{4\xi-1}(x_+ + x_-))
\label{22}
\end{eqnarray}
in the maximally extended case.
The metric (\ref{22}) may be obtained
from (\ref{23}) by choosing $f_+ = x_+$ and $f_- = A/(K^4 x_-)$, and
absorbing the $K\sim M$ factor into $h_-$.  Performing this operation on
(\ref{23}) when $\xi =1/4$ yields
\begin{eqnarray}
ds^2 &=& -2 \frac{dx_+ dx_-}{\ell(x_+ + x_-)} \nonumber\\
\phi(x) &=&  \frac{1}{2a}\ln(\frac{2A}{\ell}(x_+ + x_-))
\label{22b}
\end{eqnarray}
or
\begin{eqnarray}
ds^2 &=&
\frac{A}{\ell^2}e^{-2\ell x} dt^2 + \ell^2\frac{dx^2}{-A e^{-2\ell x}}
\nonumber\\
\phi(x) &=&  \frac{\ell}{a} x   \label{25y}\\
\psi(x) &=&  2\ell x +\psi_0 \nonumber
\end{eqnarray}
in the static case.  The solution (\ref{25y}) may be obtained from
(\ref{29}) by rescaling $x\to \ell x/M$, $t \to M t/\ell$ and then setting
$M = 0$.

For the solution (\ref{30}) there is no rescaling of $x$ that permits one
to take the $M\to 0$ limit in a non-singular way.  It is possible to
rescale the constant $x_0$ so that the $\lim_{M\to 0}(\alpha(x)) \sim
x$; the resultant spacetime is flat. However this limit is singular in that
the dilaton field is shifted by an infinite constant.

\section{Black Hole Solutions}

In this section the circumstances under which the spacetimes associated
with the solutions (\ref{25}), (\ref{29}), and (\ref{30}) correspond to
black holes will be investigated.  This will be done by requiring these
solutions to satisfy the following conditions.

\bigskip

\noindent
(1) The spacetime must be asymptotically flat, {\it i.e.} the curvature
scalar $R$ should vanish as $x\to \infty$.  This condition could be relaxed
to allow $R$ to approach a constant if one wished to consider
cosmological black holes, {\it i.e.} black hole spacetimes which are
asymptotic to (anti-)de Sitter space at large $x$.

\noindent
(2) There must be an event horizon for a finite real value of $x$. This
will occur provided $\alpha(x_H) = 0$ for some $x=x_H$, $|x_H| < \infty$.

\noindent
(3) The metric signature must be $(-,+)$ for large $x$; otherwise the
horizon is a cosmological horizon.

\noindent
(4) The ADM-mass must be real and positive. In a two dimensional spacetime
that has a timelike Killing vector $\xi^\mu$ the current
$\frac{\delta S}{\delta g^{\mu\nu}} \xi^\nu
\equiv {\cal T}_{\mu\nu}\xi^\nu$ (where $S$ is given by (\ref{1})) is
covariantly conserved. In two dimensions this yields a mass function
$\grad_\mu \hat{{\cal M}} = -\epsilon_{\mu\nu}T^\nu_\rho \xi^\rho$
from which one obtains
\be
{\cal M} = (\frac{\gamma}{4a} -\frac{1}{8\pi G})\alpha^\prime
+(\gamma+\frac{b}{a})\alpha\phi^\prime -\int\left[\frac{1}{32\pi G}
\left(\frac{(\alpha^\prime)^2-K^2}{\alpha}\right)
-\frac{b}{2}\alpha(\phi^\prime)^2
-\frac{\gamma}{2}\alpha^\prime\phi^\prime \right]  \label{36}
\ee
as the ADM-mass \cite{RtDil}. The requirement then is
that ${\cal M} > 0$. Note that $M\sim {\cal M}=0$ spacetimes are not flat;
rather they are spacetimes in which the stress-energies
of the gravitational and Liouville fields cancel each other.

\bigskip

\noindent
({\bf A}) $\alpha(x) = 2Mx - Bx_0^2 (\frac{x}{x_0})^p$

For this solution the Ricci scalar is
\be
R = -\frac{d^2}{dx^2}\alpha = B p(p-1) (\frac{x}{x_0})^{p-2} \label{32}
\ee
which, as $x \to \infty$,
vanishes for $p<2$ and approaches a constant for $p=2$. The spacetime is
singular at the origin $x=0$. An event
horizon is located at $x_H = \frac{M}{2B} e^{2a\phi_H}$, yielding from
condition (2) $\frac{B}{2M} > 0$; $\phi_H$ is the value of the Liouville
field at the horizon.  Condition (3) implies that $B$ and $M$
are both positive for $p<1$, and are both negative when $1<p<2$.
Condition (4) implies that
\be
{\cal M} = \frac{M}{8\pi G}[p-2 -(2p-3)4\pi G\frac{\gamma}{a}]
= M\frac{8\pi G\gamma(\pi G{b}+ 4\pi G a\gamma)+ a^3 -10\pi G\gamma a^2}
{4\pi G a(a^2 + 4\pi G{b})}  \label{32a}
\ee
is positive. Writing
\be
\tilde{a} = \frac{a}{\pi G |\gamma|} \qquad
\tilde{b} = \frac{b}{\pi G |\gamma|^2} \label{33}
\ee
and putting all the requirements together yields
\be
\frac{\sgn(\Lambda)}{M(4\tilde{b}-\tilde{a}^2 + 8s\tilde{a})(\tilde{b}+4)}
> 0 \qquad M\frac{8s\tilde{b}+ \tilde{a}^3 +32\tilde{a} -10 s\tilde{a}^2}
{\tilde{a}(\tilde{a}^2 + 4\tilde{b})} > 0
\nonumber
\ee
\be
1< 8\frac{\tilde{b}+ s\tilde{a}}{\tilde{a}^2 + 4\tilde{b}}< 2, \quad M < 0
\qquad
8\frac{\tilde{b}+ s\tilde{a}}{\tilde{a}^2 + 4\tilde{b}}< 1, \quad M > 0
\label{34}
\ee
provided $\gamma\neq 0$; here $s\equiv |\gamma|/\gamma$. If $\gamma = 0$
then (\ref{34}) becomes
\begin{eqnarray}
\frac{\sgn(\Lambda)}{M(4\tilde{b}-\tilde{a}^2)\tilde{b}} > 0 &\qquad &
\frac{M}
{\tilde{a}(\tilde{a}^2 + 4\tilde{b})} > 0  \label{35}\\
1< 8\frac{\tilde{b}}{\tilde{a}^2 + 4\tilde{b}}< 2, \quad M < 0
&\quad\mbox{or}\quad&
8\frac{\tilde{b}}{\tilde{a}^2 + 4\tilde{b}}< 1, \quad M > 0
\nonumber
\end{eqnarray}
where $\tilde{a}$ and $\tilde{b}$ are given by (\ref{33}) with $\gamma =
1$. Several distinct cases arise for the allowed regions of parameter
space.
\medskip

\noindent
(i) $M>0$, $\gamma > 0$. For $\Lambda>0$, (\ref{34}) becomes
\be
-\frac{\tilde{a}^2}{4} < \tilde{b} < -4  \quad\mbox{and}\quad
\frac{10\tilde{a}^2-\tilde{a}^3-32\tilde{a}}{8} < \tilde{b}
\label{36a}
\ee
for $\tilde{a}>0$; for $\tilde{a}<0$ only the first inequality applies.
For $\Lambda <0$, the allowed regions are
\begin{equation}
-4 < \tilde{b} < \frac{\tilde{a}^2-8\tilde{a}}{4}
 \quad\mbox{and}\quad  -\frac{\tilde{a}^2}{4} < \tilde{b}
\label{37}
\end{equation}
in the $\tilde{a}$--$\tilde{b}$ plane.
\medskip

\noindent
(ii) $M<0$, $\gamma > 0$. For $\Lambda >0$ the allowed regions are
\be
\frac{10\tilde{a}^2-\tilde{a}^3-32\tilde{a}}{8} < \tilde{b}
< \frac{\tilde{a}^2-8\tilde{a}}{4}  \quad\mbox{and}\quad \tilde{b}>-4,
\qquad 0 < \tilde{a} < 4 \label{38}
\ee
whereas for $\Lambda<0$,
\be
\frac{10\tilde{a}^2-\tilde{a}^3-32\tilde{a}}{8} < \tilde{b}
\quad\mbox{and}\quad \tilde{a} < 0  \label{39}
\ee
\medskip

The allowed regions of parameter space for the solution (\ref{25}) are
shown in Figures 1 ($M>0$) and 2 ($M<0$) for $s=1$ (positive $\gamma$);
the $s=-1$ case is easily obtained by reflecting
$\tilde{a}\to -\tilde{a}$.
\medskip

\noindent
(iii) $\gamma=0$. Only the $p<1$ region is allowed, implying
$M > 0$ always. $\Lambda$ and $b$ are of opposite sign, and
$a^2 > 4\pi G |b|$.
\medskip

\noindent
In each of the above cases, if one adds the additional requirement
that the Liouville field have positive kinetic energy, then all
regions with $\tilde{b} >0 $ are excluded.
\bigskip

\noindent
({\bf B}) $\alpha(x) = 1 - C e^{-2Mx}$

The solution (\ref{29}), where $b= -\frac{a^2}{4\pi G}$
is somewhat simpler to analyze. The Ricci scalar is
\be
R = \frac{8\pi G\Lambda a}{a+ 4\pi G\gamma}e^{-2M(x-x_0)}
\label{39a}
\ee
and diverges for large negative $x$; the spacetime is
asymptotically flat provided $M>0$. In this case
\be
{\cal M}  = \frac{M}{8\pi G}(1- 8\pi G\gamma/a) \label{39b}
\ee
and the horizon is at
\be
x_H = \frac{1}{2M}\ln(C) + x_0 = \frac{1}{2M}\ln\left(
\frac{2\pi G\Lambda a}{M^2(4\pi G\gamma +a)}\right) + x_0  \label{39c}
\ee
so conditions (1)--(4) give
\be
M > 0 \qquad  \frac{a \Lambda}{a+ 4\pi G\gamma} > 0
\qquad 1 > 8\pi G\frac{\gamma}{a} \label{40}
\ee
which in terms of $\tilde{a}$ as given by (\ref{33}) implies
$-4 <\tilde{a} < 0$ for $\Lambda < 0$ and $\tilde{a} < -
4$ or $\tilde{a}>8$ for $\Lambda >0$. In all cases the Liouville field
$\phi$ has positive kinetic energy.
\bigskip

\noindent
({\bf C}) $\alpha(x) = 2Mx\ln(2D x x_0)$

For the solution (\ref{30}), $b= \frac{a^2}{4\pi G} - 2a\gamma$. The Ricci
scalar is
\be
R= -2M/x \label{40a}
\ee
and diverges as $x\to 0$. The spacetime is asymptotically
flat,
\be
{\cal M} = \frac{M}{8\pi G}(1- 4\pi G\gamma/a) \label{40b}
\ee
and $x_H=\frac{M(4\pi G\gamma -a)}{4\pi G\Lambda a}e^{2a\phi_H}$,
where $\phi_H$ is the value of the Liouville field at the horizon.
Conditions (1)--(4) yield
\be
M > 0  \qquad \frac{4\pi G\gamma-a}{a \Lambda} > 0 \qquad
1 > 4\pi G\frac{\gamma}{a}
\label{41}
\ee
and consequently $\Lambda$ and $a$ are of opposite sign. The Liouville
field has positive kinetic energy only if $a > 8\pi G\gamma >0$ or
$a < 8\pi G\gamma <0$.

\section{Thermodynamic Properties}

The temperature of the black hole solutions (\ref{25}), (\ref{29}),
and (\ref{30})  may easily be computed by Euclideanizing them ($t=i\tau$)
and then requiring that the periodicity of $\tau$ be such that no conical
singularities are present. For a metric of the form (\ref{24}), this yields
a temperature \cite{MST}
\be
T = \frac{1}{4\pi} \left. |\frac{d\alpha}{dx}|\right|_{x_H} \label{42}
\ee
where $x_H$ is the location of the horizon in the coordinate system
(\ref{24}), {\it i.e.} $\alpha(x_H)=0$.

For the solutions (\ref{25}), (\ref{29}), and (\ref{30})  this yields
\be
\begin{array}{ll}
\alpha = 2Mx - Bx_0^2(\frac{x}{x_0})^p   & T = \frac{|M(1-p)|}{2\pi} \\
\alpha = 1 - Ce^{-2Mx} & T = {M}/{2\pi} \\
\alpha = 2Mx\ln(2D x x_0) & T = {M}/{2\pi}
\end{array}
\label{43}
\ee

These results may be confirmed using semiclassical methods which relate the
trace anomaly to Hawking radiation \cite{CF}. Writing
the metric in the form
\begin{equation}
ds^2=-C(u,v)du\,dv \label{29c}
\end{equation}
yields
\begin{equation}
\langle O\vert T^c_{\mu\nu}\vert O\rangle =\Theta _{\mu\nu}- c_M\hbar{R\over
48\pi}g_{\mu\nu} \label{29e}
\end{equation}
for the expectation value of the stress-energy tensor $T^c_{\mu\nu}$
for conformally coupled matter with central charge $c_M$, where
\begin{equation}
\Theta_{uu}=-{1\over 12\pi}\sqrt{C}\,\partial ^2_u\, C^{-1/2}~~~~~~~
\Theta_{vv}=-{1\over 12\pi}\sqrt{C}\,\partial ^2_v\, C^{-1/2}~~~~~~~
\Theta_{uv}=0~~~~.
\label{eq29f}
\end{equation}
Evaluating these for the three types of black hole solutions found above
yields in the $c_M=1$ case
\be
\begin{array}{ll}
\alpha = 2Mx - Bx_0^2(\frac{x}{x_0})^p  \quad p<1
& \Theta_{uu} = \Theta_{vv} =
-\frac{1}{12\pi}\left[(M(1-p))^2- \frac{4M^2p(2-3p)}{\frac{B}{2M}+
e^{-2M(1-p)(u-v)}}\right] \\
\alpha = 2Mx - Bx_0^2(\frac{x}{x_0})^p
\quad 1<p<2  & \Theta_{uu} = \Theta_{vv} =
-\frac{1}{12\pi}\left[(M(1-p))^2- \frac{4M^2p(p-2)}{e^{-2M(p-1)(u-v)}-1}
\right] \\
\alpha = 1 - Ce^{-2Mx} & \Theta_{uu} = \Theta_{vv} =
-\frac{M^2}{12\pi}\left[1-\frac{1}{1+Ce^{-2M(u-v)}}\right] \\
\alpha = 2Mx\ln(2D x x_0) & \Theta_{uu} = \Theta_{vv} =
-\frac{M^2}{12\pi} \left[ 1+ e^{4M(u-v)} \right]
\end{array}
\label{44}
\ee
and at ${\cal I}^+$ ($v\to \infty$) these each approach a constant value.
Extracting $\langle T_{tt}\rangle$ for each case in the usual manner and
recalling that the energy density of $(1+1)$ dimensional
radiation is $\pi T^2/6$, gives temperatures for each of these three cases
consistent with (\ref{43}).

Note that in each case the black hole radiation temperature is dependent
upon the ADM-mass ${\cal M}$, in contrast to the situation in string-
inspired dilaton theories of gravity \cite{Harvrev,symbh,ross2dcoll}
in which the
temperature is independent of this quantity. The entropy may be calculated
using the thermodynamic relation $d{\cal M} = T dS$. In all cases the
entropy varies logarithmically with the mass, a feature of
$(1+1)$-dimensional gravity pointed out previously \cite{TomRobb}. Hence
Liouville black holes have a positive specific heat (since the temperature
decreases with the mass) and it is possible for several such black holes to
have a larger entropy than one large black hole whose total mass is the sum
of the smaller masses, quite unlike the situation in $(3+1)$ dimensions.

The detailed circumstances under which this can occur have been
investigated in reference \cite{Ptland}. As an example consider two
systems, one consisting of a black hole of mass ${\cal M}$, and the other
consisting of two black holes of mass ${\cal M}$ and ${\cal M}-m$. The
difference in entropy between these two systems is
\begin{eqnarray}
\Delta S &=& \hat{K}\left[\ln(\frac{{\cal M}}{{\cal M}_0})
- \ln(\frac{{\cal M}-m}{{\cal M}_0}) - \ln(\frac{m}{{\cal M}_0})
\right]\nonumber \\
&=& \hat{K} \ln\left(\frac{{\cal M}}{m}\frac{{\cal M}_0}{{\cal M}-m}\right)
\label{45}
\end{eqnarray}
where ${\cal M} \ge m \ge {\cal M}_0$ and $\hat{K}$ is a constant whose value
changes depending upon which of the three solutions in (\ref{43}) is under
consideration. The two-hole system will have a smaller
entropy than the one-hole system provided
\be
{\cal M} < \frac{m^2}{m-{\cal M}_0} \qquad .\label{46}
\ee
The constant ${\cal M}_0$ is a constant of integration which corresponds to
the minimal mass a black hole system may have; it is a
fundamental gravitational length scale somewhat analogous to the
Planck mass \cite{TomRobb}.

Results for each of the solutions are given in the table
below.
\vfil\eject

\centerline{\bf Table I}
\medskip
\noindent
\hspace*{-1.0in}
\begin{tabular}{||l|c|c|c||}\hline
&({\bf A})&({\bf B})&({\bf C})\\
\hline\hline
$\alpha(x)$ & $2Mx - Bx_0^2(\frac{x}{x_0})^p $
& $1 - Ce^{-2M(x-x_0)}$ & $2Mx\ln(2D x x_0)$ \\
\hline
$\phi$ & $\frac{2-p}{2a}\ln(x/x_0) + \phi_0$ & $\frac{M}{a}(x-x_0) + \phi_0 $
& $\frac{1}{2a}\ln(\frac{4D}{M}x) $\\
\hline
$\psi$ & $-p \ln(2Mx) +\psi_0$ & $2Mx+\psi_0$ & $-2M\ln(\frac{D}{M}x)
+\psi_0$\\
\hline
Parameters& $p = \frac{4\xi}{4\xi -1} = 8\pi G \frac{a\gamma+b}{a^2+4\pi
Gb} $ & $b=-\frac{a^2}{4\pi G}$ & $b=\frac{a^2}{4\pi G}-2a\gamma$ \\
& $B = \Lambda \frac{(a^2+4\pi G b)^2}
{(b+4\pi G\gamma^2)(4\pi Gb+8\pi Ga\gamma - a^2)}$&
$C \equiv \frac{2\pi G\Lambda a}{M^2(a+ 4\pi G\gamma)}$ &
$D = \pi G\Lambda\frac{a}{4\pi G \gamma - a}$\\
\hline
${\cal M}$&$M\frac{8\pi G\gamma(\pi G{b}+ 4\pi G a\gamma)+ a^3 -10\pi G\gamma
a^2}
{4\pi G a(a^2 + 4\pi G{b})} $ & $\frac{M}{8\pi G}(1- 8\pi G\gamma/a)$
& $\frac{M}{8\pi G}(1- 4\pi G\gamma/a)$ \\
\hline
& If $M>0\qquad\qquad$ &  &  \\
Black hole & $-\frac{\tilde{a}^2}{4}\tilde{b}< -4$ & $M>0$ always
& $M>0$ always   \\
Criteria  & $10\tilde{a}^2 - \tilde{a}^3 +32\tilde{a} < 8\tilde{b}$&
$\tilde{a} >8$ or  & $0 > a > 4\pi G\gamma$ \\
($\Lambda>0$)&-----------------------------& $\tilde{a} <-4$ & \\
($\gamma > 0$) & If $M<0 \qquad\qquad$ & &\\
 & $0<\tilde{a}<4$ \& $\tilde{b}>4$ & &\\
 & $10\tilde{a}^2 - \tilde{a}^3 +32\tilde{a} < 8\tilde{b} < 2\tilde{a}^2-
 16\tilde{a}$& &\\
\hline
& If $M>0 \qquad\qquad$ &  &  \\
Black hole & $-\frac{\tilde{a}^2}{4}\tilde{b}$ and &  &   \\
Criteria  & $-4< \tilde{b}< \frac{\tilde{a}^2-8\tilde{a}}{4}$ &
$M>0$ always & $M>0$ always \\
($\Lambda<0$)&-----------------------------& $-4 < \tilde{a} < 0$ &
$a > 4\pi G\gamma > 0$\\
($\gamma > 0$) & If $M<0 \qquad\qquad$ & &\\
 & $\tilde{a}<0$ and & &\\
 & $10\tilde{a}^2 - \tilde{a}^3 +32\tilde{a} < 8\tilde{b}$ & &\\
\hline
Horizon & $x_H= \frac{{\cal M}}{\Lambda}\zeta_1$ &
$x_H = \frac{a-8\pi G\gamma}{16\pi G{\cal M}a} \ln(
\frac{\Lambda}{(4\pi G{\cal M})^2}\zeta_2)$
& $x_H = -\frac{2{\cal M}}{\Lambda}e^{2a\phi_H}$ \\
\hline
Temperature &$2\hat{K}{\cal M}$ & $4G{\cal M}\frac{a}{a-8\pi G\gamma}$
  &$4G{\cal M}\frac{a}{a-4\pi G\gamma}$\\
\hline
Entropy& $\frac{1}{2\hat{K}}
\ln({\cal M}/{\cal M}_0)$
& $\frac{a-8\pi G\gamma}{4Ga}\ln({\cal M}/{\cal M}_0)$
& $\frac{a-4\pi G\gamma}{4Ga}\ln({\cal M}/{\cal M}_0)$ \\
\hline
\end{tabular}
\bigskip
\hspace*{0in}

\noindent
where in the table
$$
\hat{K} \equiv \frac{a(8\pi G a\gamma+4\pi G b- a^2)} {a^3-10\pi G
\gamma a^2 + 32 (\pi G\gamma)^2 a + 8 (\pi G)^2 b \gamma}
$$
$$
\zeta_1 \equiv
4\pi G \hat{K}e^{2a\phi_H}\frac{b+4\pi G\gamma^2}{a^2+4\pi G b}
\qquad
\zeta_2 \equiv \frac{(a-8\pi G\gamma)^2 e^2}{4a(4\pi G\gamma+a)}
$$
\vspace{1.0in}

{}From table I one can construct the following scenario for radiating
Liouville black holes. As the hole radiates its mass ${\cal M}$ decreases
at a rate
\be
\frac{d{\cal M}}{dt} = -\frac{\pi}{6} T^2 = - \kappa {\cal M}^2\label{46a}
\ee
where $\kappa$ is a function of the coupling constants $a$, $b$ and
$\gamma$ that depends upon which of the solutions  ({\bf A})--({\bf C}) is
under consideration. This implies that a black hole of initial mass
${\cal M}_i$ decays at a rate
\be
{\cal M}(t) = \frac{{\cal M}_i}{1+\kappa{\cal M}_i t} \label{46b}
\ee
lowering its temperature and entropy.  Superficially it appears that the
black hole is long-lived, since it takes infinitely long for the mass to
radiate completely away.  However at a finite time $t_f=({\cal M}_i/{\cal
M}_0 -1)/(\kappa{\cal M}_i)$ the mass approaches the constant ${\cal M}_0$
at which point the entropy vanishes and this semi-classical picture no
longer applies.

For solutions ({\bf A}) and ({\bf C}) the horizon moves closer to the
singularity at $x=0$; this (proper) distance eventually becomes comparable
to the Compton wavelength of a body of mass ${\cal M}$,
after which the solution breaks down   and
quantum effects due to the Liouville field can no longer be ignored.
Setting ${\cal M}_0$ to be the mass at which the extremal
proper length from the horizon to the singularity is equal to the Compton
wavelength of a particle of mass ${\cal M}_0$, {\it i.e.}
\be
\int_0^{x_H}\frac{dx}{\sqrt{|\alpha(x)|}} = \frac{\hbar}{{\cal M}_0}
\label{46c}
\ee
yields
\begin{eqnarray}
{\cal M}_0 &=& \sqrt{G\hbar\left(\frac{\hbar\Lambda}{\hat{\zeta}}
\right)}\frac{\Gamma\left(\frac{p}{2(p-1)}\right)}{\Gamma\left(
\frac{2p-1}{2(p-1)}\right)} \qquad \mbox{for ({\bf A})} \\
{\cal M}_0 &=& \sqrt{G\hbar\left(\frac{4\hbar|\Lambda|a}
{(a-4\pi G\gamma)}\right)} e^{-a\phi_H}
\qquad \mbox{for ({\bf C})} \label{46d}
\end{eqnarray}
where
\be
\hat{\zeta} = \frac{(8\pi G\gamma a + 4\pi G b -a^2)(\pi G b+ (2\pi
G\gamma)^2)}{(a^2 + 4\pi G b)^2}e^{2a\phi_H} \label{46e}
\ee
and the dependence on $\hbar$ has explicitly been included.
Note that $G\hbar$ is dimensionless and units have been chosen so that
the speed of light is unity.

Solution ({\bf B}) has qualitatively different behaviour. For a given value
of $x_0$, $x_H$ is minimized at
\be
{\cal M}_{m} = e \frac{a-8\pi G\gamma}{8\pi G a}\sqrt{
\frac{2\pi G a}{4\pi G\gamma +a}} \label{46f}
\ee
where $e=2.71828...$
As the hole radiates, the mass ${\cal M}$ will decrease to this
value, the location of the horizon also decreasing from its initial value.
The extremal proper length varies inversely with $M$ and so
(\ref{46c}) implies
\be
a(1- 32 G\hbar) \ge 8\pi G\gamma \label{46g}
\ee
which provides another constraint on the coupling parameters $a$ and
$\gamma$, but does not provide a value for ${\cal M}_0$.
As ${\cal M}$ falls below ${\cal M}_{m}$, the location of the
horizon then begins to increase without bound, and the black hole comes to
dominate the entire spacetime. Presumably quantum gravitational
effects come
into play before this takes place, a condition which can be realized
by requiring that
the Compton wavelength of a body of mass ${\cal M}$ always be smaller than
the proper length from the horizon to the singularity for a body of
mass ${\cal M}_{m}$ so that
\be
\int_0^{x_H({\cal M}_m)}\frac{dx}{\sqrt{|\alpha(x)|}} \ge
 \frac{\hbar}{{\cal M}}
\label{46h}
\ee
or
\be
{\cal M} \ge
\frac{e\hbar}{\pi}\sqrt{\frac{2\pi G\Lambda a}{4\pi G\gamma + a}}
\equiv {\cal M}_0
\label{46i}
\ee
which permits a definition of ${\cal M}_0$.

\section{Quantum Corrections}

The conservation of $\langle O|T^c_{\mu\nu}|O\rangle$ combined with the trace
anomaly
\be
\langle O\vert T_{\mu}^{c\ \mu}\vert O\rangle
= -\frac{c_M\hbar}{24\pi}R\label{47}
\ee
(which follows from (\ref{29e})) yields
\begin{eqnarray}
\langle O|T^c_{+-}| O\rangle &=& \frac{1}{12\pi}\partial_+\partial_-\rho
\nonumber \\
\langle O|T^c_{\pm\pm}|O\rangle &=& \frac{1}{12\pi}\left(
(\partial_\pm\rho)^2 \partial^2_\mu\rho +\hat{t}_\pm(x_\pm) \right)
\label{48}
\end{eqnarray}
where $\hat{t}_\pm(x_\pm)$ depend upon the choice of boundary conditions.

Hence incorporation of the conformal stress-energy to one-loop order
modifies (\ref{12} -- \ref{15}) to
\be
\dif_{+}\dif_{-}(\psi + 2\rho) = 0            \label{49}
\ee
\be
\dif_{+}\dif_{-}\rho = \frac{\pi G\Lambda(a\gamma+b)}{b(1-\frac{c_M\hbar
G}{3})
+4\pi G\gamma^2}
e^{2(\rho-a\phi)} \label{50}
\ee
\be
\dif_{+}\dif_{-}\phi = \frac{\Lambda(a(1-\frac{c_M\hbar G}{3})
-4\pi G\gamma)}{4(b(1-\frac{c_M\hbar G}{3})+4\pi G\gamma^2)}
e^{2(\rho-a\phi)} \label{51}
\ee
\begin{eqnarray}
\frac{1}{2}(\dif_{\pm}\psi)^2 -\dif^2_\pm\psi
+ 2 \dif_\pm\rho\dif_\pm\psi
+ 8\pi G(b(\dif_{\pm}\phi)^2 -\gamma\dif^2_\pm\phi
+ 2\gamma \dif_\pm\rho\dif_\pm\phi)
&& \nonumber\\
\quad + 2\frac{c_M\hbar G}{3}\left((\partial_\pm\rho)^2 \partial^2_\mu\rho
+\hat{t}_\pm \right) + T^c_{\pm\pm}  = 0&&
\label{52}
\end{eqnarray}
where $T^c_{\mu\nu}$ is the classical part of the stress-energy of the
conformal matter. There is also an equation of motion for the conformal
matter fields; for example, if there are $N$ scalar fields $\{{\cal F}_I\}$,
then
\be
T^c_{\mu\nu} = \frac{1}{8\pi}\sum_{I=1}^N
\left(\grad_\mu{\cal F}_I \grad_\nu{\cal F}_I -
\frac{1}{2}g_{\mu\nu}(\grad{\cal F}_I)^2\right) \label{52b}
\ee
and
\be
\dif_{+}\dif_{-}{\cal F}_I  = 0 \label{52a}
\ee
are the equations of motion of the ${\cal F}$--fields, valid to one-loop.
Solutions to equations (\ref{49}--\ref{52}) incorporate the one-loop
back-reaction of the conformal stress-energy into the spacetime geometry.
The classical stress-energy tensor $T^c_{\mu\nu}$ has no effect on
the evolution of the metric/Liouville system; rather it modifies the
evolution of the dilaton field $\psi$ via (\ref{52}).  Its quantum
corrections (from (\ref{47})) are easily deduced:
a brief inspection of this system of equations indicates that (\ref{49}--
\ref{52}) are respectively {\it identical} to (\ref{12}--\ref{15}) but with
the gravitational constant $G$ renormalized to
\be
G_R = \frac{G}{1- \frac{c_M\hbar G}{3}} \label{53}
\ee
and the $\psi$-field renormalized to
\be
\psi_R = \frac{1}{1-\frac{c_M\hbar G}{3}}\psi +2\rho\frac{c_M\hbar
G}{3-c_M\hbar G} \quad .
\label{53a}
\ee

Hence quantum corrections due to conformally coupled matter map the metric
\be ds^2 = -\alpha(x;G) dt^2 +  \frac{dx^2}{\alpha(x;G)}     \label{54}
\ee
to
\be ds^2 = -\alpha(x;G_R) dt^2 + \frac{dx^2}{\alpha(x;G_R)}
\label{55}
\ee
an expression which fully takes into account the back-reaction
to one-loop order.  As a consequence, the
solutions (\ref{25}), (\ref{29}) and (\ref{30}) are robust: the parameters
$p$, $B$, $D$ and ${\cal M}$ become functions of $c_M$, and a solution with
a given set of these parameters is mapped into solution with another set of
these parameters.

Specifically, for the solution (\ref{25}) (case ({\bf A}))
\begin{eqnarray}
p_R &=& 8\pi G\frac{a\gamma+b}{a^2(1-\frac{c_M\hbar G}{3})+4\pi G b}
\label{56c}
\label{56}\\
x_{H_R} &=& \frac{M}{2\Lambda} \frac{(b(1-\frac{c_M\hbar G}{3})+4\pi G\gamma^2)
(4\pi Gb+8\pi Ga\gamma - a^2(1-\frac{c_M\hbar G}{3}))e^{2a\phi_H}}
{(a^2(1-\frac{c_M\hbar G}{3})+4\pi G b)^2 } \nonumber \\
{\cal M}_R &=& M\frac{8\pi G\gamma(\pi G{b}+ 4\pi G a\gamma)+ a^3
(1-\frac{c_M\hbar G}{3})^2 -10\pi G\gamma a^2  (1-\frac{c_M\hbar G}{3})}
{4\pi G a(a^2 (1-\frac{c_M\hbar G}{3}) + 4\pi G{b})} \nonumber
\end{eqnarray}
and so (for Liouville fields with $b<0$) the parameter $p$ increases,
decreasing the parameter range for solutions obeying the flatness
condition. Although $x_H$ increases as $c_M\hbar \le 3/G$ increases, the
extremal
proper length from the singularity to the horizon (slowly)
decreases with increasing $c_M\hbar < 3/G$; back-reaction effects `shrink' the
size of the black hole.

For the solution (\ref{29}) (case ({\bf B}))
(which now holds if $b = -\frac{a^2}{4\pi G}(1-\frac{c_M\hbar G}{3})$)
\begin{eqnarray}
x_{H_R} &=& \frac{1}{2M}\ln \left(\frac{2\pi G\Lambda a}{M^2 (4\pi G\gamma
+ a (1-\frac{c_M\hbar G}{3}) )} \right) + x_0 \nonumber \\
{\cal M}_R  &=&  \frac{M}{8\pi G a} (a(1-\frac{c_M\hbar G}{3}) - 8\pi G\gamma)
\label{56a}
\end{eqnarray}
and so the horizon moves outward and the mass decreases provided
$a>4\pi G\gamma$. There is no effect on the extremal proper length
from the singularity to the horizon as it varies like $1/M$.
Note that the mass goes negative for
$c_M\hbar  > 3/G(1-\frac{8\pi G\gamma}{a})$.

Finally for case ({\bf C}) (eq. ({\ref{30})),
for which $b = \frac{a^2}{4\pi G}(1-\frac{c_M\hbar G}{3}) - 2a\gamma$  (and
$\Lambda < 0$)
\begin{eqnarray}
x_{H_R} &=& \frac{M}{2\pi G |\Lambda| a} \left( a(1-\frac{c_M\hbar G}{3})
- 4\pi G\gamma\right) e^{2a\phi_H}    \nonumber \\
{\cal M}_R &=& \frac{M}{8\pi G} \left( 1-\frac{c_M\hbar G}{3}
- \frac{4\pi G\gamma}{a}\right)  \label{56b}
\end{eqnarray}
and the effects of the back-reaction in this case are to decrease the
location of both the horizon and the mass. The extremal proper length
(which here varies as $\sqrt{x_H}$)  therefore also decreases for positive
mass black holes.

The case $c_M\hbar G = 3$ may also be treated, provided that $\gamma\neq 0$.
In this case
\begin{eqnarray}
p_R &=& 2(1+a\gamma/b) \nonumber\\
x_{H_R} &=& \gamma^2 \frac{2a\gamma+b}{\Lambda b^2}   \label{57}\\
{\cal M}_R &=& M\frac{\gamma(b+4a\gamma)}{2ab}
\end{eqnarray}
In this case it is straightforward to show that
asymptotically flat black hole solutions of positive mass exist provided
\begin{eqnarray}
p<1 \quad &\mbox{and}& \quad M>0 \nonumber\\
\mbox{If}\quad \Lambda > 0 : \qquad  &a\gamma < 0& 4|a\gamma| > b > 2 |a\gamma|
\label{58}\\
\mbox{If}\quad \Lambda < 0 : \qquad &a\gamma > 0 & -4 a\gamma > b \nonumber\\
                    & a\gamma < 0 & -2 a\gamma > b \nonumber
\end{eqnarray}
or
\begin{eqnarray}
1<p<2 \quad &\mbox{and}& \quad M<0 \nonumber\\
\mbox{If} \Lambda < 0  & b< 0& 4 a\gamma > |b| > 2 a\gamma
\label{58a}\\
\mbox{If} \Lambda > 0  &a\gamma > 0 & b > -2 a\gamma  \nonumber\\
                    & a\gamma < 0 & b> -4 a\gamma  \nonumber \quad .
\end{eqnarray}

One can include quantum gravitational effects by  considering
a functional integral over the field configurations of the metric, $\psi$,
and $\phi$ fields \cite{2dqgm}. The path integral is
\be
Z=\int \frac{{\cal D}g}{V_{GC}}
{\cal D}\psi {\cal D}\phi {\cal D}\Phi e^{-(S[\psi,g,\phi] +S_M[\Phi])\hbar}
\label{25q}
\ee
where $S=S[\psi,g,\phi]+S_M[\Phi]$ is the Euclideanization of the action
(\ref{5}), with $S_M$ being the part of the action incorporating
additional matter fields $\{\Phi\}$. The volume of the diffeomorphism group,
$V_{GC}$, has been factored out.

Making the same scaling assumption as in refs. \cite{2dqgm,DDK} about
the functional measure  yields
\begin{eqnarray}
Z&=&\int [{\cal D}\tau] {\cal D}_g\phi {{\cal D}_g b} {{\cal D}_g c}
{\cal D}_g\psi {\cal D}_g\phi{\cal D}_g\Phi
e^{-(S[\psi,\rho,\phi]+S_{gh}[b,c]+S_M[\Phi])/\hbar}
\label{26q}\\
&=&\int [{\cal D}\tau] {\cal D}_{\hat{g}}\phi {{\cal D}_{\hat{g}} b}
{{\cal D}_{\hat{g}} c} {\cal D}_{\hat{g}}\psi {\cal D}_{\hat{g}}\phi
{\cal D}_{\hat{g}}\Phi
e^{-(S[\psi,\rho,\phi]+S_{gh}[b,c]+S_M[\Phi]+\hat{S}[\rho,\hat{g}])/\hbar}
\nonumber
\end{eqnarray}
where $[{\cal D}\tau]$ represents the integration over the Teichmuller
parameters and
\be
\hat{S}[\rho,\hat{g}]=\frac{\hbar}{8\pi}\int d^2x \sqrt{\hat{g}}\left(
\hat{g}^{\mu\nu}\partial_\mu\rho\partial_\nu\rho - Q \rho\hat{R} \right)
\label{27q}
\ee
is the Liouville action with arbitrary coefficients where $g_{\mu\nu}=
e^{\beta\rho}\hat{g}_{\mu\nu}$, and $\hat{R}$  and
$\hat{\grad}^2$  are respectively the curvature scalar and Laplacian of the
metric  $\hat{g}$. $S_{gh}$ is the action for the ghost fields $b$ and $c$.

The action $S_{\rm TOT} =
\hat{S}+S[\psi,g,\phi]+S_{gh}+S_M $ may be rewritten as
\begin{eqnarray}
S_{\rm TOT} &=&  S_{gh} + S_M +
\frac{1}{8\pi G} \int d^2x \sqrt{\hat{g}}\left[
-\frac{1}{2}\hat{g}^{\mu\nu}\partial_\mu\psi\partial_\nu\psi
- 8\pi G b\hat{g}^{\mu\nu}\partial_\mu\phi\partial_\nu\phi \right.
\label{59} \\
 & - & \left.
(\psi+8\pi G\gamma\phi) (\hat{R} - \beta \hat{\grad}^2\rho)
+ G\hbar (\hat{g}^{\mu\nu}\partial_\mu\rho\partial_\nu\rho - Q \rho\hat{R})
-8\pi G \Lambda)e^{\beta\rho -2a\phi} \right] \nonumber
\end{eqnarray}
where the cosmological constant has been renormalized to zero and the
coupling $e^{-2a\phi}$ has been gravitationally dressed.

The approach is then to determine the parameters $\beta$ and $Q$ from the
requirement that the conformal anomaly vanish and that $e^{\beta\rho-
2a\phi}$ is a conformal tensor of weight (1,1).
Upon rescaling the fields $\psi$, $\rho$ and $\phi$ so that
\begin{eqnarray}
\tilde{\rho}&=&\sqrt{\hbar+\frac{\beta^2}{2G} + 2\pi\frac{(\beta\gamma)^2}{b}}
\rho \nonumber \\
\tilde{\psi}&=&\frac{1}{\sqrt{2G}}\left(\psi+\beta\phi\right) \label{60}\\
\tilde{\phi}&=& \sqrt{8\pi G |b|}(\phi+\frac{\beta\gamma}{2b}\rho)
\nonumber
\end{eqnarray}
(\ref{59}) becomes
\begin{eqnarray}
S_{\rm TOT}&=& \frac{1}{8\pi}\int d^2x \sqrt{\hat{g}}\left\{
\hat{g}^{\mu\nu}\partial_\mu\tilde{\rho}\partial_\nu\tilde{\rho}
-\hat{g}^{\mu\nu}\partial_\mu\tilde{\psi}\partial_\nu\tilde{\psi}
-\sgn(b)\hat{g}^{\mu\nu}\partial_\mu\tilde{\phi}\partial_\nu\tilde{\phi}
\right. \nonumber \\
&& \left. \qquad - \sqrt{\frac{2}{G}}\tilde{\psi}\hat{R}
- \sqrt{\frac{8\pi\gamma^2}{|b|}}\tilde{\phi}\hat{R}
- \frac{\hbar Q-\frac{\beta}{G} - 4\pi \gamma^2\frac{\beta}{b}}
{\sqrt{\hbar+\frac{\beta^2}{2G} + 2\pi\frac{(\beta\gamma)^2}{b}}}\trho\hat{R}
\right. \nonumber \\
&&\left. -8\pi\Lambda
\exp\left[\frac{\beta(1-\frac{a\gamma}{b})}{\sqrt{1+\frac{\beta^2}{2G}}}
\trho\right] -\frac{2a}{\sqrt{8\pi|b|}}
\right\} + S_{gh} + S_M \quad . \label{61}
\end{eqnarray}
The coefficients in front of the $\tilde{\psi}$ and $\tphi$
terms are due to the signs of the kinetic energy terms in
(\ref{5}). They may be dealt with by rescaling $\tpsi\to i\tpsi$
and, if $b>0$, $\tphi$ to $i\tphi$ so that the functional integral converges.

The action in (\ref{61}) may now be analyzed using conformal field
theoretic techniques.  The fields $\tphi$, $\tpsi$ and $\trho$
have propagators
\be
<\tphi(z)\tphi(w)> = - \ln(z-w) = <\tpsi(z)\tpsi(w)>  \label{62}
\ee
and the contribution of the matter and ghost fields is as usual.
Hence it is straightforward to compute the total central charge
\begin{eqnarray}
c_{\rm TOT} &=& 1 + 3 \frac{(Q-\frac{\beta}{G\hbar}
- 4\pi \gamma^2\frac{\beta}{b\hbar})^2}
{1+\frac{\beta^2}{2G\hbar} + 2\pi\frac{(\beta\gamma)^2}{b\hbar}} + 1
-\frac{6}{G\hbar} +1 -24\frac{\pi\gamma^2}{b\hbar} -26 + c_M
\nonumber \\
&=& c_M-23 + 3\frac{\hat{G}\hbar Q^2-2(1+\beta Q)}{\hat{G}\hbar+\beta^2/2}
\label{63}
\end{eqnarray}
where $\hat{G}\equiv \frac{b G}{b+4\pi G\gamma^2}$.
The requirement that $e^{\beta\rho-
2a\phi}$ be a conformal tensor of weight (1,1) yields
\be
-\frac{1}{2}\frac{\beta \Upsilon}{1+\frac{\beta^2}{2\hat{G}\hbar}}
( Q-\frac{\beta}{\hat{G}\hbar} + \beta\Upsilon ) +\Delta_\phi -1  = 0
\label{64}
\ee
where $\Upsilon \equiv 1 - \frac{a\gamma}{2b}$ and
\be
\Delta_\phi = \frac{1}{2}\frac{a}{8\pi b}(8\pi\gamma + a\hbar)
\label{65}
\ee
is the conformal dimension of the Liouville field.

Equation (\ref{64}) yields the constraint
\be
Q^2 \ge
8(1-\Delta_\phi)(1+ \frac{1-\Delta_\phi -\Upsilon}{\Upsilon^2\hat{G}\hbar})
\label{66}
\ee
since $\beta$ must be real. Setting $c_{\rm TOT}=0$ and using
(\ref{64}) gives
\be
\beta^2 = \frac{6\Upsilon-12 +\hat{G}\hbar(\Upsilon(23-c_M)-12\Sigma) \pm
\sqrt{\Omega}}
{\Upsilon(c_M-23+6\hat{G}\hbar\Sigma^2+12\Sigma)}
\label{67}
\ee
where
\be
\Omega \equiv (\Upsilon^2((c_M-23)\hat{G}\hbar -6)
+ 24 \Upsilon(1+\hat{G}\hbar\Sigma)-24) ((c_M-23)\hat{G}\hbar-6)
\label{67z}
\ee
and $\Sigma \equiv \Upsilon -
\frac{1-\Delta_\phi -\Upsilon}{\Upsilon\hat{G}\hbar}$.

Equation (\ref{67})
determines $\beta = \beta(G;c_M)$ in terms of $c_M$, $G$ and the
coupling constants $a$, $b$ and $\gamma$. As $\beta$ must be real and
positive, this will put constraints on $a$, $b$ and $\gamma$ in
terms of $c_M$ and $G$. Choosing $\hat{g}_{+-} = -\frac{1}{2}$
(analogous to (\ref{11})) one finds that the classical equations of
motion (\ref{12} -- \ref{15}) become the same as (\ref{49} -- \ref{52})
with the replacements
\be
\rho\rightarrow \hat{\rho} = \frac{2}{\beta}\rho \quad \mbox{and} \quad
c_M \rightarrow -\frac{12}{\beta^2} \label{68}
\ee
and so quantum gravitational effects map the classical metric (\ref{54})
\be
ds^2 = e^{2\rho(G)}dx_+ dx_- \label{69a}
\ee
to
\be
ds^2 = e^{2\beta\rho(G_R)}dx_+ dx_ = e^{2\hat{\rho}(G_R)}dx_+ dx_-
\label{69}
\ee
where
\be
G_R = \frac{G}{1 + \frac{4 G\hbar}{\beta^2}} \label{70}
\ee
is now the renormalized gravitational constant.

\section{Discussion}

Liouville fields coupled to two-dimensional gravity provide a rich
theoretical laboratory for investigating the puzzles of black
hole radiation.  In the present paper the most general set of
exact solutions to the field equations associated with the action
(\ref{1}) have been obtained; a wide class of these solutions
correspond to asymptotically flat two-dimensional black holes of
positive mass. In contrast to the situation for the
string-theoretic black hole \cite{Callan}, the field equations yield an
exact solution (to one-loop order) to the back-reaction problem
for both conformally coupled matter and for the situation in
which both the metric and Liouville field are quantized. In each
case quantum effects map a solution with coupling
constant $G$ to another solution with $G\to G_R$.

The above considerations suggest the following scenario for the evaporation
of Liouville black holes. For definiteness, consider the situation
when $b=-\frac{a^2}{4\pi G}$ and $\gamma=0$ \cite{Qlou}.
When quantum effects are taken into account, the
classical spacetime described by (\ref{29}) is modified to
that described by (\ref{25}) where
\begin{eqnarray}
\alpha_Q &=& 2\tilde{M}x -
\frac{4\pi G\Lambda}{(1+\frac{\beta^2}{4G\hbar})(1+\frac{\beta^2}{2G\hbar})}
x_0^2(\frac{x}{x_0})^{-\frac{\beta^2}{2G\hbar}} \nonumber\\
\phi_Q &=& (1+\frac{\beta^2}{4G\hbar})\ln(\frac{x}{x_0})
\quad\mbox{and}\quad  \psi_B = \frac{\beta^2}{2G\hbar}\ln(\frac{x}{x_0})
+\psi_0 \label{71}
\end{eqnarray}
where $\beta$ is given by (\ref{67}),
$\tilde{M} = \frac{M}{1+\beta^2/(2G\hbar)}$ and
the subscript ``Q'' denotes the fact that the fully quantized
back-reaction has been included.
The reality condition on $\beta$ yields, from
(\ref{67})
\be
6 > G\hbar(c_M-23) > 6(1-(G\hbar)^2) \label{73}
\ee
thereby constraining the range of values of the matter central charge in
terms of the dimensionless coupling $G\hbar$.

The spacetime described by the metric in (\ref{71}) is that of an
asymptotically flat positive mass black hole for large positive $x$.
It has
a singularity in the curvature, dilaton and Liouville fields at $x=0$.
The entropy is
\be
S = \frac{(1+4\frac{G\hbar}{\beta^2})^2}{1+2\frac{G\hbar}{\beta^2}}
\ln\left(\frac{{\cal{M}}_Q}{{\cal{M}}_0}\right)
\label{74}
\ee
and the temperature $T=\frac{M}{2\pi}$.

The ADM-mass is now
\be
{\cal M}_Q = \frac{M}{8\pi G}
\frac{(1+4\frac{G\hbar}{\beta^2})^2}{1+2\frac{G\hbar}{\beta^2}}
\label{72}
\ee
and is manifestly positive. The inclusion of the quantum stress-energy
results in a larger ADM-mass.  This mass
will decrease with time as ${\cal M}_Q \sim 1/t$,
decreasing both the location
$x_H$ of the horizon and the maximal proper distance from the horizon
to the singularity. Eventually this distance (the ``size'' of the
event horizon) becomes comparable to the Compton wavelength as discussed
in section 5. Since  as ${\cal M}_Q \to {\cal M}_0$ the
entropy tends to zero for finite temperature, at this point the
thermodynamic description breaks down and higher-loop corrections become
important.

What effect might these have? Previous work \cite{MST} suggests that
quantum vacuum energies  $\Lambda_Q$ (renormalized to zero to this order)
will modify the temperature  to ${T} \sim \sqrt{M^2 -
\frac{\Lambda_Q}{2}}$, so that entropy becomes
$$
S \sim \ln\left( \frac{\sqrt{M^2 - \frac{\Lambda_Q}{2}} +
M}{\tilde{M}_0}\right)
$$
where $\tilde{M}_0$ is a constant comparable in magnitude to $M_0$. In this
case the black hole slowly cools off to a zero temperature remnant, leaving
behind a global event horizon with its requisite loss of quantum coherence.
Whether or not  higher-loop effects will have similar consequences is
a subject for further investigation.

\section*{Acknowledgements}
This work was supported by the Natural Sciences and Engineering Research
Council of Canada.

\section*{Figure Captions}

\noindent
{\bf Fig. 1} Shaded areas correspond to the allowed regions in
($\tilde{a}$,$\tilde{b}$) parameter space for $M>0$.

\noindent
{\bf Fig. 2} Shaded areas correspond to the allowed regions in
($\tilde{a}$,$\tilde{b}$) parameter space for $M<0$.

 \end{document}